\documentclass{article}
\usepackage{mlspconf}
\usepackage{amsmath}
\usepackage{amssymb}
\usepackage{graphicx}
\graphicspath{{./}{pics/}}

\copyrightnotice{\vspace*{12pt}\footnotesize\parbox{\textwidth}{\copyright\ 2013 IEEE. This is the authors' postprint version of the article. The original print version appeared as: Tuomo Sipola, Fengyu Cong, Tapani Ristaniemi, Vinoo Alluri, Petri Toiviainen, Elvira Brattico, and Asoke K. Nandi. Diffusion map for clustering fMRI spatial maps extracted by independent component analysis. In Machine Learning for Signal Processing (MLSP), 2013 IEEE International Workshop on, Southampton, United Kingdom, September 2013. IEEE.}}

\title{Diffusion map for clustering fMRI spatial maps extracted by independent component analysis}

\newcommand{\jyumit}{\textsuperscript{1}}
\newcommand{\jyumus}{\textsuperscript{2}}
\newcommand{\uk}{\textsuperscript{3}}
\newcommand{\hel}{\textsuperscript{4}}
\newcommand{\aal}{\textsuperscript{5}}
\newcommand{\upcom}{\textsuperscript{,}}

\name{
\begin{tabular}{c}
Tuomo Sipola\jyumit\upcom\sthanks{Tuomo Sipola's work was supported by the Foundation of the Nokia Corporation and the Finnish Foundation for Technology Promotion.}, 
Fengyu Cong\jyumit\upcom\sthanks{This work was partially supported by TEKES (Finland) grant 40334/10 ‘Machine Learning for Future Music and Learning Technologies’.}, 
Tapani Ristaniemi\jyumit, 
Vinoo Alluri\jyumit\upcom\jyumus, \\ 
Petri Toiviainen\jyumus, 
Elvira Brattico\jyumus\upcom\hel\upcom\aal,
Asoke K. Nandi\uk\upcom\jyumit\upcom\sthanks{Asoke K. Nandi would like to thank TEKES for the award of the Finland Distinguished Professorship.}
\end{tabular}
}
\address{
\jyumit Department of Mathematical Information Technology, \\ University of Jyv\"{a}skyl\"{a}, Finland \\
\jyumus Finnish Centre of Excellence in Interdisciplinary Music Research, \\ University of Jyv\"{a}skyl\"{a}, Finland \\
\uk Department of Electronic and Computer Engineering, \\ Brunel University, United Kingdom \\
\hel Cognitive Brain Research Unit, Institute of Behavioral Sciences, \\  University of Helsinki, Finland \\
\aal Brain \& Mind Laboratory, Biomedical Engineering and Computational Science, \\ Aalto University, Espoo, Finland
}

\begin{document}

\maketitle

\begin{abstract}
Functional magnetic resonance imaging (fMRI) produces data about activity inside the brain, from which spatial maps can be extracted by independent component analysis (ICA).  
In datasets, there are $n$ spatial maps that contain $p$ voxels. The number of voxels is very high compared to the number of analyzed spatial maps. 
Clustering of the spatial maps is usually based on correlation matrices. This usually works well, although such a similarity matrix inherently can explain only a certain amount of the total variance contained in the high-dimensional data where $n$ is relatively small but $p$ is large. 
For high-dimensional space, it is reasonable to perform dimensionality reduction before clustering. In this research, we used the recently developed diffusion map for dimensionality reduction in conjunction with spectral clustering. 
This research revealed that the diffusion map based clustering worked as well as the more traditional methods, and produced more compact clusters when needed. 
\end{abstract}


\section{\MakeUppercase{Introduction}}
\label{sec:intro}

In order to gain understanding about the human brain, various technologies have recently been introduced, such as electroencephalography (EEG), tomography, magnetoencephalography (MEG) and functional magnetic resonance imaging (fMRI). They provide scientists with data about the temporal and spatial activity inside the brain. Functional magnetic resonance imaging is a brain imaging method that measures blood oxygenation level. It detects changes in this level, that are believed to be related to neurotransmitter activity. This enables the study of brain functioning, pathological trait detection and treatment response monitoring. The method localises brain function well, and thus is useful in detecting differences in subject brain responses \cite{MATTHEWS2004, HUETTEL2009}. 

Deeper understanding about the simultaneous activities in the brain begins with a decomposition of the data. Independent component analysis (ICA) has been extensively used to analyze fMRI data. It tries to decompose the data into multiple components that are mixed in the original data. Basically, there are two ways to perform ICA: group ICA and individual ICA \cite{CALHOUN2009}. Group ICA is performed on the data matrix including all the participants' fMRI data, and individual ICA is applied on each dataset of each participant. Among datasets of different participants, group ICA tends to need more assumptions which are not required by individual ICA \cite{CONG2013}. 
For individual ICA, if the components for each participant are known, it is expected to find the most common components among the participants. 
Therefore, clustering spatial maps extracted by ICA is a necessary step for the individual ICA approach to find common spatial information across different participants in fMRI research. 

ICA decomposes the individual datasets and creates components that can be presented with spatial maps. After ICA has been applied, a data matrix of size $n$ by  $p$ is produced, where $n$ is the number of spatial maps and $p$ is the number of voxels of each spatial map. The $n$ spatial maps come from different participants, and $n$ is much smaller than $p$ in fMRI research. Clustering the spatial maps is mostly done using the $n$ by $n$ similarity matrix of the $n$ by $p$ data matrix \cite{CALHOUN2009, ESPOSITO2005}. Surprisingly, it usually works well although such a similarity matrix inherently can just explain a certain amount of the total variance contained in the high-dimensional $n$ by $p$ data matrix \cite{ESPOSITO2005}.

New mathematical approaches for functional brain data analysis should take into account the characteristics of the data analyzed. As stated, spatial maps have high dimensionality $p$. In machine learning, dimensionality reduction is usually performed on such datasets before clustering. In the small-$n$-large-$p$ clustering problem, the conventional dimensionality reduction methods, for example, principal component analysis (PCA) \cite{JOLLIFFE2002}, might not be suitable for the non-linear properties of the data. In this research, we apply a recently developed non-linear method called diffusion map \cite{COIFMAN2006a,NADLER2006} for dimensionality reduction. The probabilistic background of the diffusion distance metric will give an alternative angle to this dataset by facilitating the clustering task and providing visualization. This paper explores the possibility of using the diffusion map approach for fMRI ICA component clustering. 

\section{\MakeUppercase{Methodology}}
\label{sec:method}

This paper considers a dimensionality reduction approach to clustering of high-dimensional data. The clustering procedure flows as follows:

\begin{enumerate}
\item Data normalization with logarithm
\item Neighborhood estimation
\item Dimensionality reduction with diffusion map
\item Spectral clustering
\end{enumerate}

Data normalization should be done if the features are on differing scales. This ensures that the distances between the data points are meaningful. Neighborhood estimation for diffusion map creates the neighborhood where connections between data points are considered. Dimensionality reduction creates a new set of fewer features that still retain most information. Spectral clustering groups similar points together. 

We assume that our dataset consists of vectors of real numbers: $X = \left\{ x_1, x_2, \dots , x_n \right\}, x_i \in \mathbb{R}^p$. In practice the dataset is a data matrix of size $n \times p$, whose rows represent the samples and columns the features. In this study each row vector is a spatial map and column vector contains the corresponding voxels in different spatial maps. 

\subsection{Diffusion map}

Diffusion map is a dimensionality reduction method that embeds the high-dimensional data to a low-dimensional space. It is part of the manifold learning method family and can be characterized with its use of diffusion distance as the preserved metric \cite{COIFMAN2006a}. 

The initial step of the diffusion map algorithm itself calculates the affinity matrix $W$, which has data vector distances as its elements. Here Gaussian kernel with Euclidean distance metric is used \cite{COIFMAN2006a, NADLER2008}. For $\epsilon$  selection, see below. The affinity matrix is defined as  

\begin{equation*}
W_{ij} = \exp \left( -\frac{|| x_i - x_j ||^2}{\epsilon} \right),
\label{KERNEL}
\end{equation*}

\noindent where $x_i$ is the $p$-dimensional data point. The neighborhood size parameter $\epsilon$ is determined by finding the linear region in the sum of all weights in $W$, while trying different values of $\epsilon$ \cite{COIFMAN2008,SINGER2009}. The sum is 

\begin{equation*}
L = \sum_{i=1}^{n} \sum_{j=1}^{n} W_{i,j},
\end{equation*}

From the affinity matrix $W$ the row sum diagonal matrix $D_{ii} = \sum_{j=1}^{n} W_{ij}, i \in 1 \ldots n$ is calculated. The $W$ matrix is then normalized as $P = D^{-1} W$. This matrix represents the transition probabilities between the data points, which are the samples for clustering and classification. The conjugate matrix $\tilde{P} = D^{\frac{1}{2}} P D^{-\frac{1}{2}}$ is created in order to find the eigenvalues of $P$. In practice, substituting $P$, we get

\begin{equation*}
\tilde{P} = D^{-\frac{1}{2}} W D^{-\frac{1}{2}}.
\label{NGL}
\end{equation*}

This so-called normalized graph Laplacian  \cite{CHUNG1997} preserves the eigenvalues \cite{NADLER2008}. Singular value decomposition (SVD) $\tilde{P} = U \Lambda U^*$ yields the eigenvalues $\Lambda = \mathrm{diag}([\lambda_1, \lambda_2, \dots, \lambda_n])$ and eigenvectors in matrix $U = [ u_1, u_2, \dots, u_n ]$. The eigenvalues of $P$ and $\tilde{P}$ stay the same. It is now possible to find the eigenvectors of $P$ with $V = D^{-\frac{1}{2}} U$ \cite{NADLER2008}. 

The low-dimensional coordinates in the embedded space $\Psi$ are created using $\Lambda$ and $V$: 

\begin{equation*}
\Psi = V \Lambda.
\label{MAP_COORDINATES}
\end{equation*}

Now, for each $p$-dimensional point $x_i$, there is a corresponding $d$-dimensional coordinate, where $d \ll p$. The number of selected dimensions depends on how fast the eigenvalues decay. The coordinates for a single point can be expressed as

\begin{equation}
\Psi_d : x_i \to \left[ \lambda_2 v_2(x_i), \lambda_3 v_3(x_i), \dots, \lambda_{d+1} v_{d+1}(x_i) \right].
\label{DM}
\end{equation}

The diffusion map now embeds the data points $x_i$ while preserving the diffusion distance to a certain bound given that enough eigenvalues are taken into account \cite{COIFMAN2006a}. 

\subsection{Spectral clustering}

Spectral clustering is a method to group samples into clusters by benefitting from the results of spectral methods that reveal the manifold, such as the diffusion map. Spectrum here is understood in the mathematical sense of spectrum of an operator on the matrix $P$. The main idea is that the dimensionality reduction has already simplified the clustering problem so that the clustering itself in the low-dimensional space is an easy task. This leaves the actual clustering for any clustering method that can work with real numbers \cite{NG2001, KANNAN2004, LUXBURG2007}. 

The first few dimensions from the diffusion map represent the data up to a relative precision, and thus contain most of the distance differences in the data \cite{COIFMAN2006a}. Therefore, some of the first dimensions will be used to represent the data. Threshold at $0$ in the embedded space divides the space between the possible clusters, which means that a linear classification can be used. With the linear threshold, the second eigenvector separates the data into two clusters in the low-dimensional space. This eigenvector solves the normalized cut problem, which means that there are small weights between clusters but the internal connections between the members inside the cluster are strong. Clustering in this manner happens through similarity of transition probabilities between clusters \cite{NG2001, KANNAN2004, SHI2000, MEILA2000}. 

\section{\MakeUppercase{Results}}
\label{sec:results}

The data comes from experiments where participants listened to music. The data analysis was performed on a collection of spatial maps of brain activity. After dimensionality reduction and spectral clustering, the results are presented and compared to more traditional methods. 

\subsection{Data description}

In this research the fMRI data are based on the data sets used by Alluri et al.\ \cite{ALLURI2012}. Eleven musicians listened to a 512-second modern tango music piece during the experiment. In the free-listening experiment the expectation was to find relevant brain activity significantly correlating with the music stimulus. The stimuli were represented by musical features used in music information retrieval (MIR) \cite{ALLURI2012}.   

After preprocessing, PCA and ICA were performed on each dataset of each participant, and 46 ICA components (i.e., spatial maps) were extracted for each dataset \cite{PUOLIVALI2013, TSATSHIVILI2013}. Then, temporal courses of the spatial maps were correlated with one musical feature, Brightness \cite{ALLURI2012}. As long as the correlation coefficient was significant (statistical $p$-value $<0.05$), the spatial maps were selected for further analysis. Altogether, $n=23$ spatial maps were selected from 11 participants.  The number of voxels for each spatial map was $p=209{,}633$. So, the 23 by 209,633 data matrix was used for the clustering to find the common spatial map across the 11 participants. 

\begin{figure}[tb]
\centering
\includegraphics[width=8.5cm]{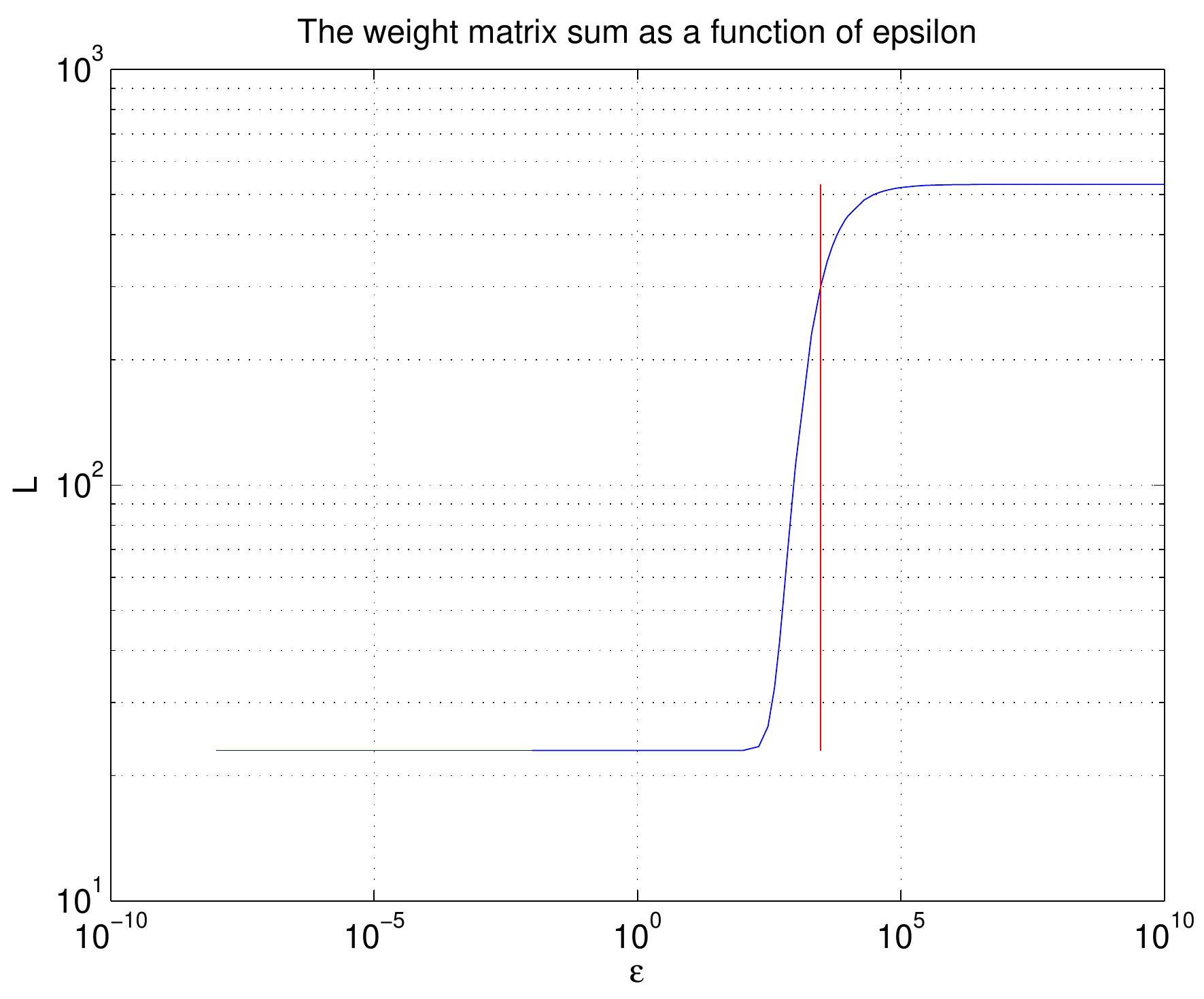}
\caption{Selecting $\epsilon$ for diffusion map. The red line shows the selected value.}
\label{fig:epsilon}
\end{figure}

\subsection{Data analysis}

The data matrix was analyzed using the methodology explained in Section \ref{sec:method}. The dimensionality of the dataset was reduced and then the spectral clustering was carried out. The weight matrix sum for $\epsilon$ selection is in Figure \ref{fig:epsilon}; the used value is in the middle region, highlighted with straight vertical line. Clustering was performed with only one dimension in the low-dimensional space. To compare the results with more traditional clustering methods, the high-dimensional data was clustered with agglomerative hierarchical clustering \cite{XU2005R} with Euclidean distances using the similarity matrix \cite{ESPOSITO2005} and $k$-means algorithms \cite{XU2005R}. The clustering results for two clusters were identical using all the methods. 

\begin{figure}[tb]
\centering
\includegraphics[width=8.5cm]{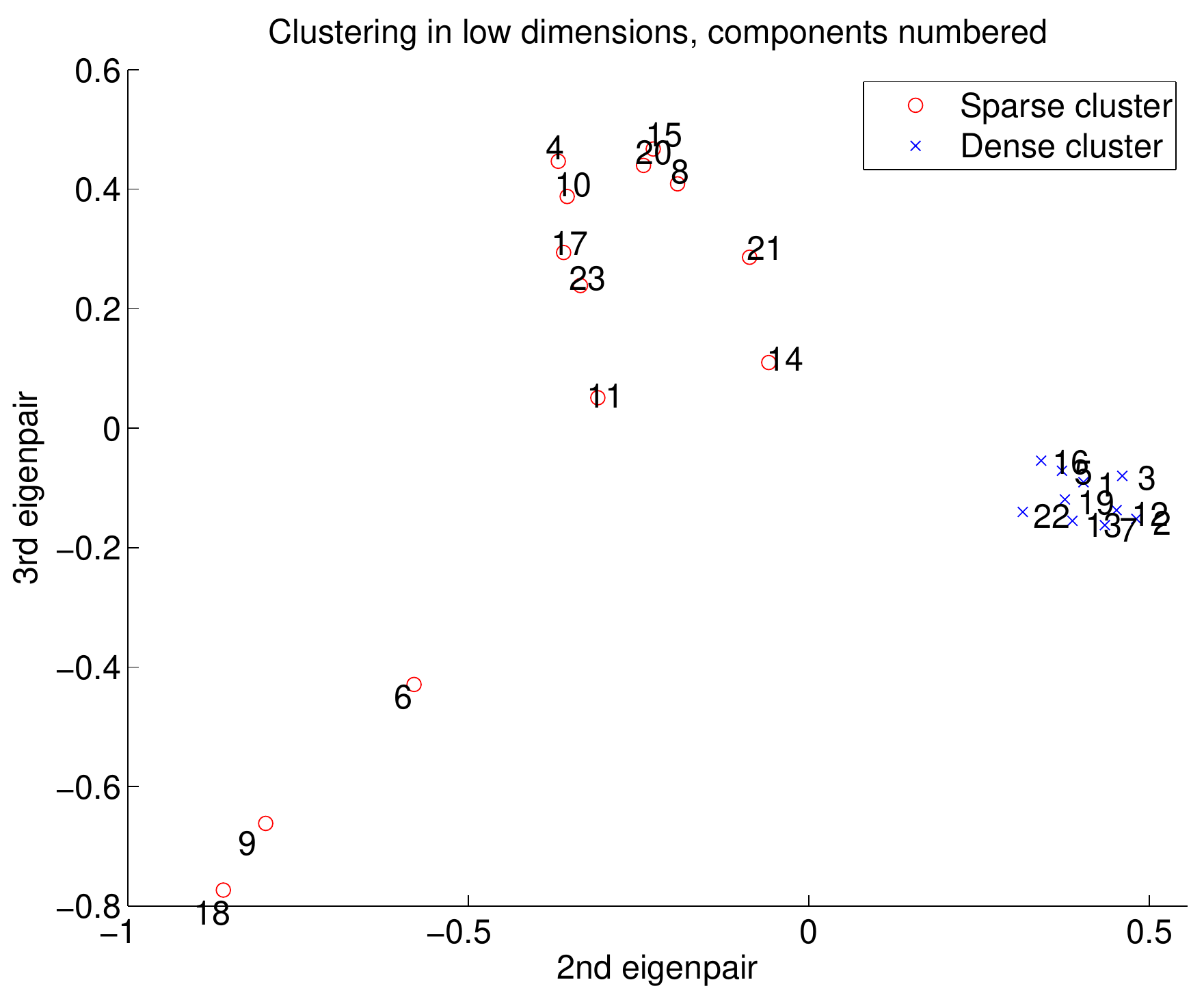}
\caption{Diffusion map clustering results.}
\label{fig:dm_clustering}
\end{figure}

\begin{figure}[htb]
\centering
\includegraphics[width=8.5cm]{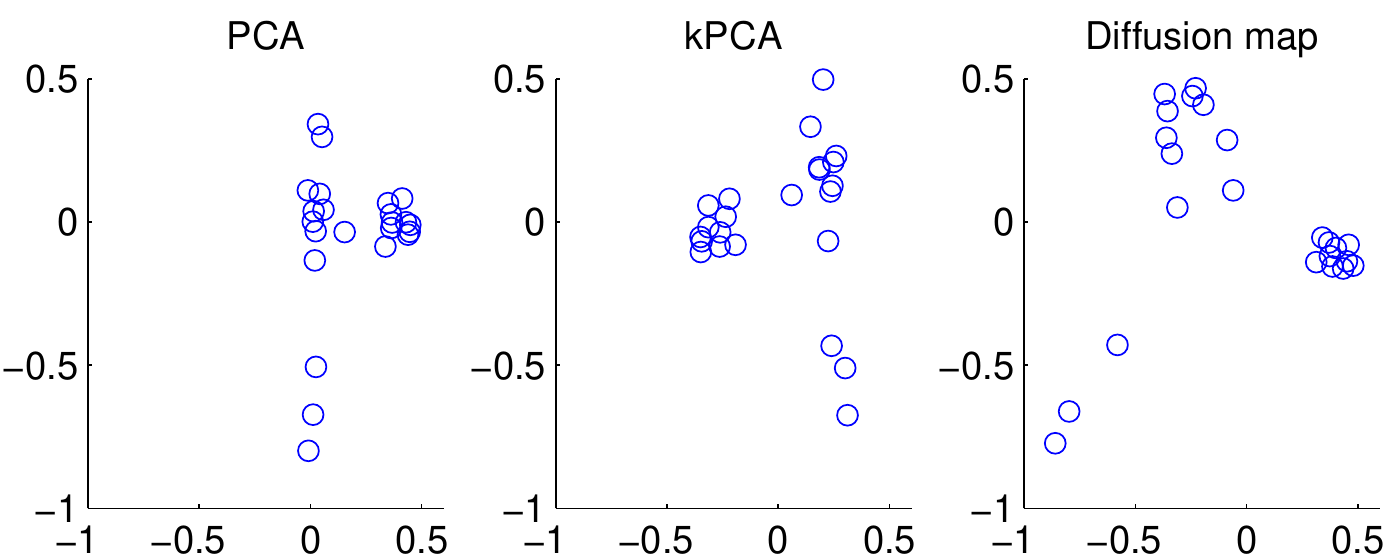}
\caption{Dimensionality reduction method comparison. The coordinates have been scaled.}
\label{fig:pca}
\end{figure}

Figure \ref{fig:dm_clustering} shows the resulting clustering from the diffusion map. The figure uses the first two eigenpairs for low-dimensional presentation, for these two clusters even one dimension is enough. The spatial maps are numbered and the two clusters are marked with different symbols. The dividing spectral clustering line is at $0$ along the horizontal axis, so the point to the right of $0$ are in one cluster and to the left another. Two clusters, dense and sparse, are detected using this threshold. The dense cluster, marked with crosses, contains components that are considered to be similar according to this clustering. The traditional PCA and kernel PCA with Gaussian kernel for spectral clustering are compared to the diffusion map \cite{MULLER2001,WANG2012}. In Figure~\ref{fig:pca} diffusion map with correct $\epsilon$ creates more firm connections, which eases the clustering task. The effect of diffusion distance metric is also seen. 

\begin{figure}[htb]
\centering
\includegraphics[width=8.5cm]{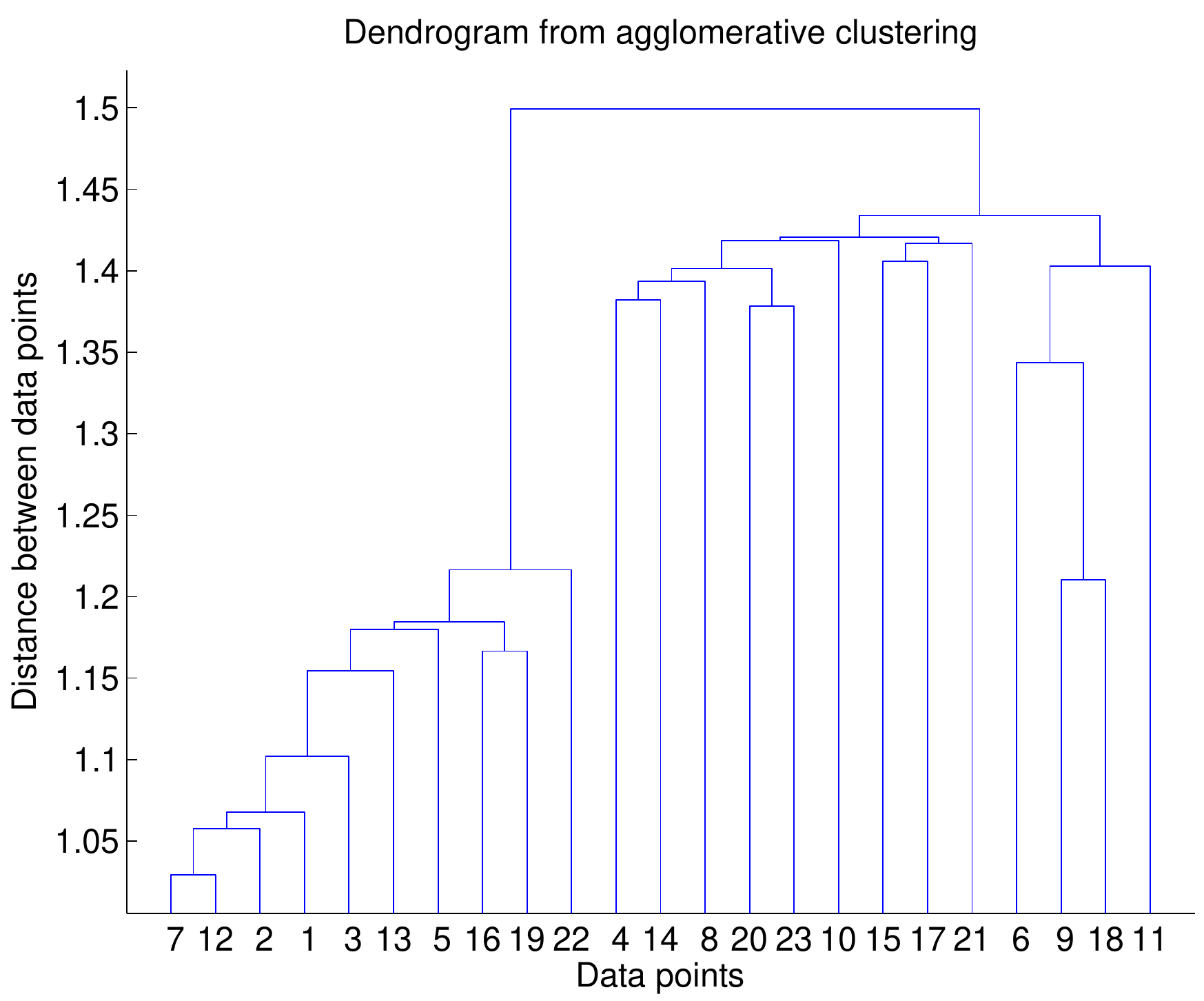}
\caption{Dendrogram from the agglomerative clustering.}
\label{fig:aggl}
\end{figure}

In Figure \ref{fig:aggl} the dendrogram produced by the agglomerative clustering is shown. The clustering results are the same as with the dimensionality reduction approach. The separation is visible at the highest level and the structure corresponds to the distances seen in Figure \ref{fig:dm_clustering}. All the points in, e.g., the dense cluster in Figure \ref{fig:dm_clustering} are in the left cluster of Figure \ref{fig:aggl}. This comparison shows the evident separation between the two clusters and also validates the results from diffusion map methodology. 

\begin{figure}[htb]
\centering
\includegraphics[width=8.5cm]{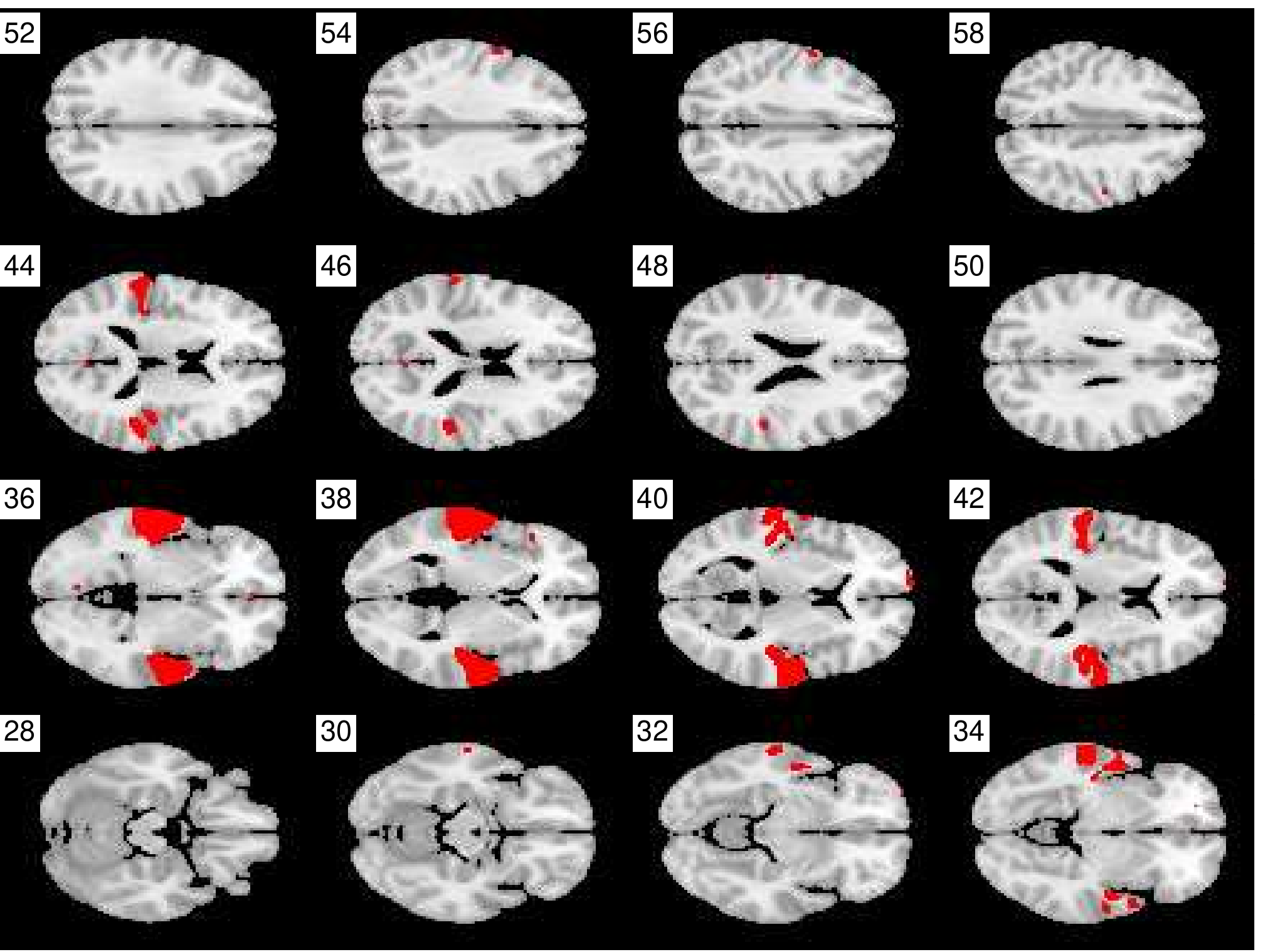}
\caption{Example spatial map in the dense cluster, this is data point number 3. Dark lateral areas mark more than three standard deviations from the mean, e.g.\ in slices 36 and 38.}
\label{fig:brain3}
\end{figure}

Figure \ref{fig:brain3} illustrates the kind of spatial maps that are found in the dense cluster. Dark areas along the lateral sides is used to highlight those voxels whose values differed more than three standard deviations from the mean. The numbers marking the slices are their Z-coordinates. The corresponding low-dimensional point is in Figure \ref{fig:dm_clustering} numbered as 3. It is now possible to inspect the clusters more closely with domain experts. 

\begin{figure}[htb]
\centering
\includegraphics[width=8.5cm]{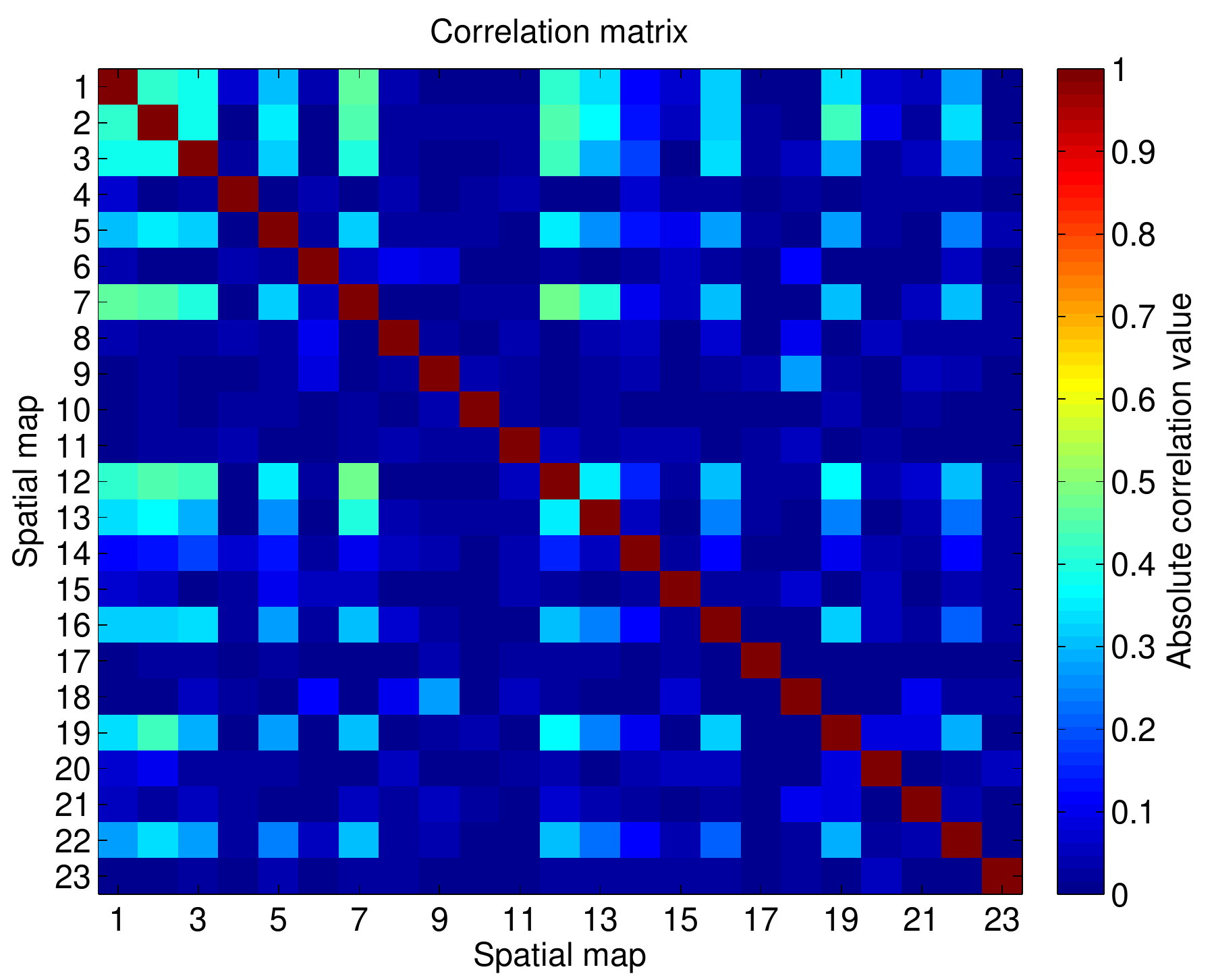}
\caption{Absolute correlation values between the spatial maps.}
\label{fig:corr_matrix}
\end{figure}

Figure \ref{fig:corr_matrix} shows the correlation matrix of all the 23 spatial maps. This is a way to inspect the similarity of the brain activity. The correlation matrix is also the basis of analysis for the hierarchical clustering \cite{ESPOSITO2005}. In the figure it can be seen that there is some correlation between some of the spatial maps, but not so much between others. 

\begin{figure}[htb]
\centering
\includegraphics[width=8cm]{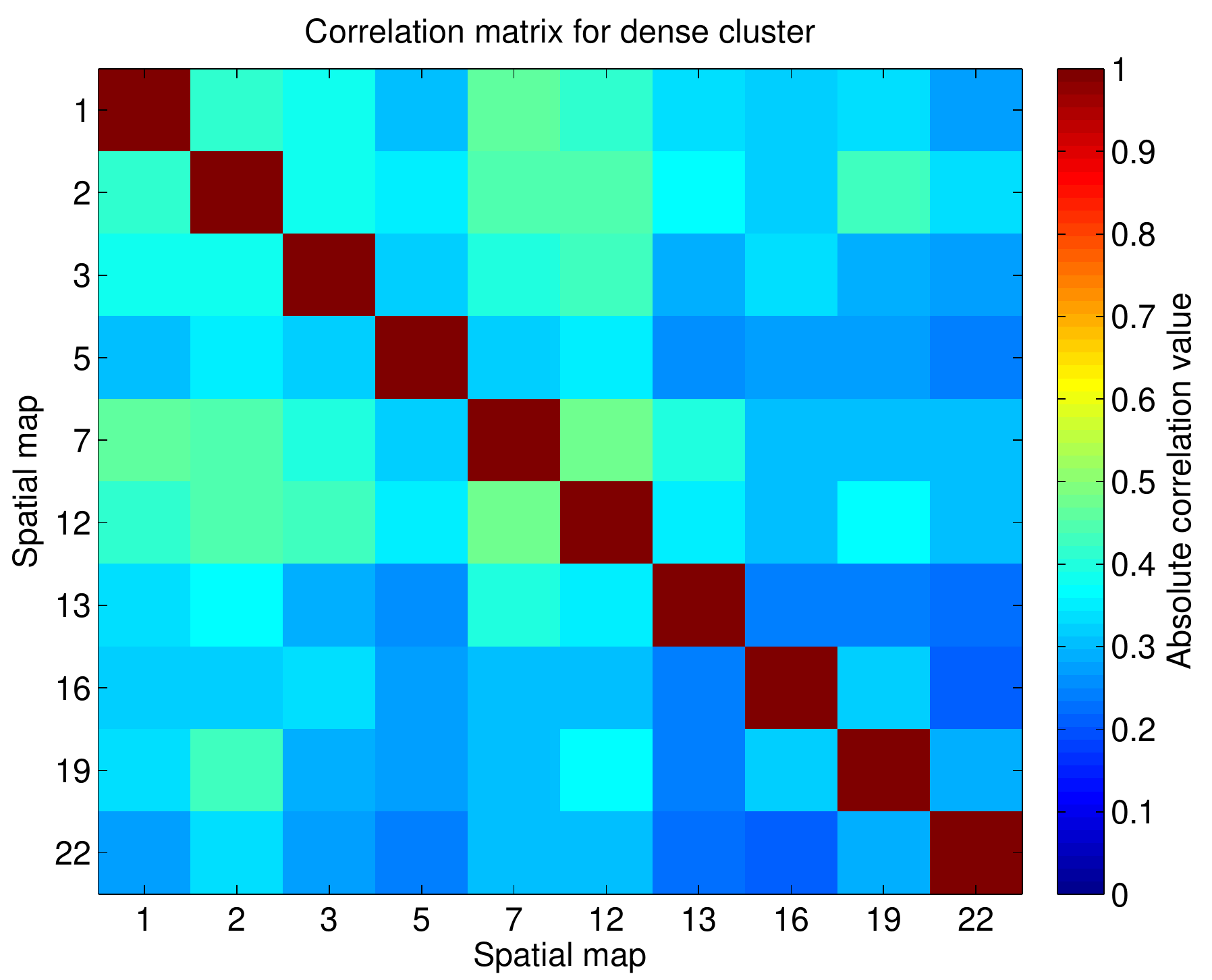}
\caption{Absolute correlation values between the spatial maps that belong to the dense cluster.}
\label{fig:corr_dense}
\end{figure}

\begin{figure}[htb]
\centering
\includegraphics[width=8cm]{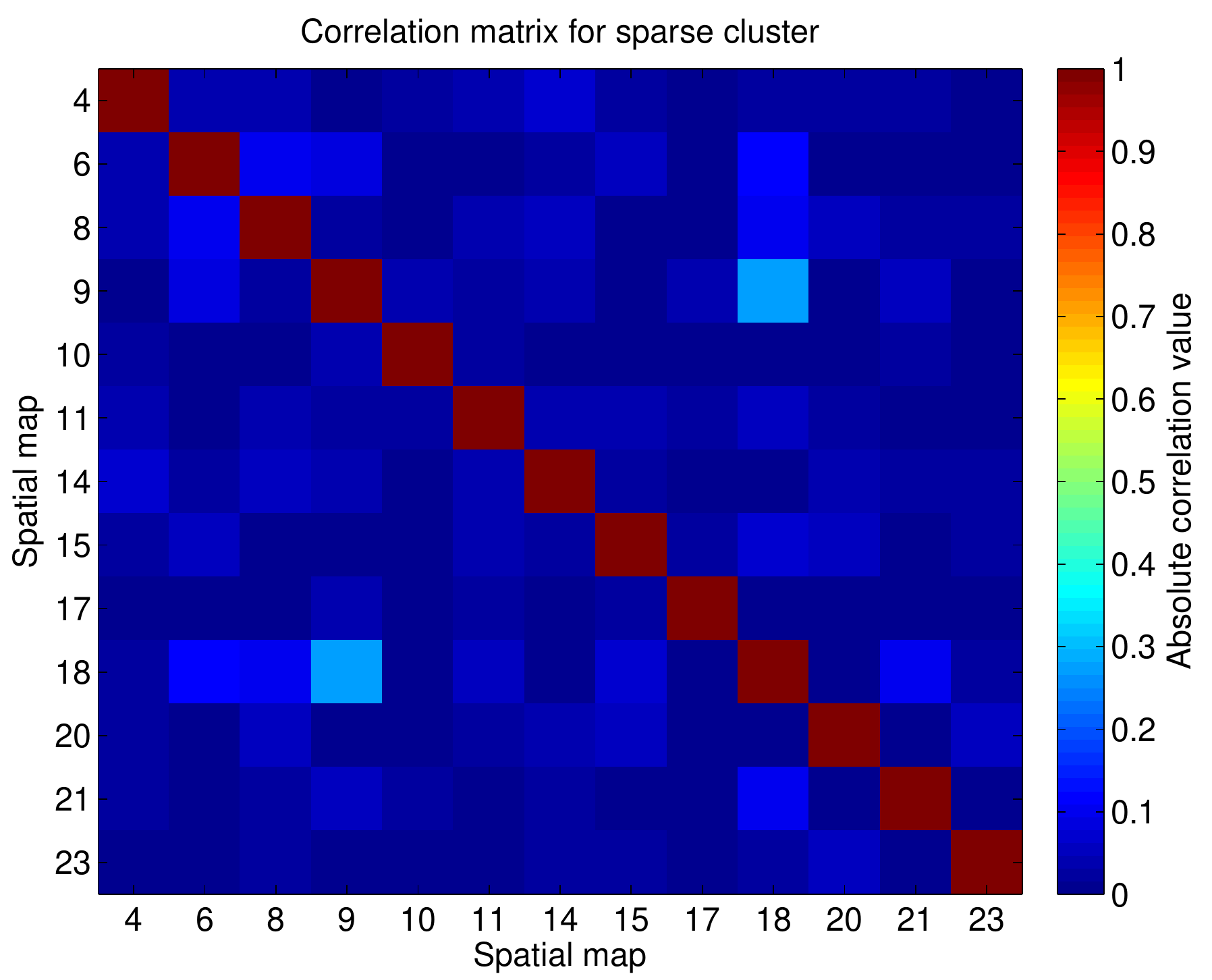}
\caption{Absolute correlation values between the spatial maps that belong to the sparse cluster.}
\label{fig:corr_sparse}
\end{figure}

Figures \ref{fig:corr_dense} and \ref{fig:corr_sparse} illustrate the internal structure of the clusters by showing the correlation matrices for the individual clusters. The members in the dense cluster have higher correlation among themselves than the members in the sparse cluster. This information is also seen in Figure \ref{fig:dm_clustering} where the diffusion distances inside the dense cluster are smaller. 

\section{\MakeUppercase{Discussion}}
\label{sec:discussion}


In this paper we have proposed a theoretically sound non-linear analysis method for clustering ICA components of fMRI imaging. The clustering is based on diffusion map manifold learning, which reduces the dimensionality of the data and enables clustering algorithms to perform their task. This approach is more suitable for high-dimensional data than just applying clustering methods that are designed for low-dimensional data. The assumption of non-linear nature of brain activity also promotes the use of methods designed for such problems. Particularly, the advantage of diffusion map is in visualizing the distribution of all data samples ($n$ spatial maps with $p$ voxels in each) by using only two coordinates. As seen in the visualization, it becomes more straightforward to determine the compact cluster from the two-dimensional plot derived from the 209,633-dimensional feature space than from the similarity matrix. 


The results show that the proposed methodology separates groups of similarly behaving spatial maps. Results from diffusion map spectral clustering are similar to hierarchical agglomerative clustering and $k$-means clustering. Small sample size and good separation of clusters makes the clustering problem rather simple to solve. 
Moreover, the visualization obtained from diffusion map offers an interpretation for clustering. 

The proposed methodology should be useful for analyzing the function of the brain and understanding which stimuli create similar spatial responses in which group of participants. The domain experts can gain more basis for the interpretation of brain activity when similar activities are already clustered using automated processes suitable for the task. Furthermore, visualization helps to identify the relationships of the clusters. 

Diffusion map execution times become increasingly larger if the number of samples goes very high. This can be overcome to a certain degree with out-of-sample extension. Big sample sizes are also a problem with traditional clustering methods. However, diffusion map offers a non-linear approach, and is suitable for high-dimensional data. Both properties are true for fMRI imaging data. 

The analysis could be expanded to more musical features and to bigger datasets in order to further validate its usefulness in understanding the human brain during listening to music. The method is not restricted only to certain kind of stimulus, so it is usable with diverse fMRI experimental setups. Furthermore, situations where traditional clustering fails when processing spatial maps, the proposed methdodology might give more reasonable results. 

\bibliographystyle{IEEEbib}

\end{document}